\documentclass[letterpaper]{article}
\usepackage[T1]{fontenc}
\usepackage{geometry}
\usepackage{setspace}
\usepackage[
  backend=biber,
  style=chem-acs,
  sorting=none
]{biblatex}

\usepackage{graphicx}
\usepackage{float}
\usepackage{amsmath}
\usepackage{amssymb}
\usepackage{subcaption}
\newfloat{scheme}{htbp}{los}
\floatname{scheme}{Scheme}
\floatname{chart}{Chart}
\newfloat{graph}{htbp}{loh}
\usepackage{chemformula}
\usepackage[version = 4]{mhchem}
\usepackage{amsmath}

\usepackage{authblk}

\author[1]{Md Mozakker H. Shojib\textsuperscript{\textdagger}}
\author[1,2]{Asier C. Monasterio\textsuperscript{\textdagger}}
\affil[1]{Institute of Biological and Chemical Systems - Functional Molecular Systems, Karlsruhe Institute of Technology, Germany}
\affil[2]{Bilbao School of Engineering, University of the Basque Country, Spain}
\author[3,4]{Emanuele Locatelli}
\affil[3]{Department of Physics and Astronomy, University of Padua, Italy}
\affil[4]{Istituto Nazionale Di Fisica Nucleare, Sezione di Padova, Italy}

\author[5,6]{\protect\\{Pascal Friederich}}
\affil[5]{Institute of Nanotechnology, Karlsruhe Institute of Technology, Germany}
\affil[6]{Institute for Anthropomatics and Robotics, Karlsruhe Institute of Technology, Germany}
\author[7]{Christopher Ness}
\affil[7]{School of Engineering, University of Edinburgh, United Kingdom}
\author[1]{Iliya D. Stoev*}

\title{\textbf{Synergistic approach to probing the dynamics and mechanics of patchy soft matter}}
\date{\textsuperscript{\textdagger}Authors with equal contribution
\bigskip 
\\
*Email: iliya.stoev@kit.edu}

\begin{document}

\maketitle

\begin{abstract}
Tailoring microscopic details to tune bulk rheology is a key paradigm in soft matter physics,
yet the vast parameter space associated with constituent interactions precludes a fully systematic approach.
To address this,
we have designed a synergistic strategy to explore the parameter space that comprises simulations,
experimental rheology,
and machine learning.
As a case study, we choose DNA-based self-assembled fluids whose viscoelastic response can be fine-tuned by manipulating the base sequencing of the constituent nucleic acid nanostars.
We use coarse-grained simulations, benchmarked against experimental data,
to obtain the rheology of the DNA fluids,
which feeds forward to a framework of Gaussian Process Regression and active learning.
The latter is then used to explore the rheological design space with high predictive precision.
The pipeline is designed to be deployed iteratively for the rational design and accelerated discovery of generic soft matter suspensions.
\end{abstract}

\section*{Keywords}

\textit{bead-spring model, structure-property relations, material properties, machine learning, in-silico rheology}

\section{Introduction}

A central goal in soft condensed matter is to establish quantitative structure-property relationships that link microscopic design variables to macroscopic material performance. This challenge is particularly prominent in self-assembling colloidal suspensions and polymeric networks; a well-known and experimentally accessible platform within this broader class are 
DNA-based bulk fluids. Here, sequence-encoded complementarity follows strict Watson-Crick base-pairing rules and enables highly selective, reversible bonding. This sequence programmability allows fine-tuning of the geometry and valency of DNA nanomotifs, which opens new opportunities for controlling gel formation, designing responsive architectures and even constructing information-carrying materials~\cite{doricchi2022emerging,biswas2017nasa,cho2015fast,de2021materials,rietsche2022quantum,can2025mechanically,gadzekpo2025integrative,meiser2020reading,Akintayo2021StarPEGDNA,Cai2025DNASupramolecularPolymerization,Pertici2020InjectableHydrogelsPerspective}. Beyond finding targeted applications in data storage and programmable matter, these systems serve as foundational model patchy networks in which multi-valency and tunable bond lifetimes can be effectively isolated and studied systematically~\cite{xing2019structural,stoev2020on}.
In general, the presence of reversible bonds and finite valency leads to a viscoelastic response that depends strongly on connectivity and relaxation dynamics~\cite{rovigatti2018simulate,stoev2020on,StaufferAharony1994,xing2019structural}. Developing predictive models that capture how interaction strength, particle geometry, stiffness and number density control the rheological fingerprint is essential for the rational design of soft materials with tailored thermodynamic and mechanical properties~\cite{gadzekpo2025integrative}.

An additional challenge is that macroscopic observables, such as melting curves
and linear viscoelastic moduli, emerge from many-body assembly pathways that are difficult to resolve or access directly in experiments. This motivates the wide use of computational models that connect microscopic design choices to bulk behavior at the scale of tens to thousands of building blocks. High-resolution nucleotide-level models (\textit{e.g.}, oxDNA) are powerful but are computationally too demanding for systematic studies of network formation and mechanics in large volumes over long times~\cite{xing2019structural,sengar2021primer,soto2025modelling}. Higher-level coarse-grained descriptions, such as based on a bead-spring model, retain only the minimal physical ingredients required for recapitulating experimental bulk behavior, thereby offering generality and universality of the obtained insights~\cite{stoev2021bulk}.

While coarse-grained simulations accelerate design predictions, an important question is whether such simplified descriptions can accurately capture experimentally observed behavior in actual DNA hydrogels. For example, experimental systems composed of Y-shaped nanostars connected by complementary linkers exhibit characteristic viscoelastic spectra and tunable mechanical responses that depend on sticky-end binding strength and structural flexibility~\cite{can2025mechanically}. Therefore, establishing a correspondence between simulation parameters and experimentally measurable rheological properties is crucial to boost the predictive power of the model, allowing it to guide experimental effort and accelerate material design. 

In this work, we introduce an efficient approach for the exploration of the parameter space of self-assembling fluids and, in general, soft matter systems, focusing specifically on the case of DNA hydrogels. This approach is based on the synergistic combination of coarse-grained modeling, Brownian dynamics simulations and machine learning (ML), with final validation against experimental data. 

First, a suitable computational model should be identified. Here, we choose to adapt a minimal bead-spring and patchy-particle description of DNA-based Y-shaped building blocks; this model has already been employed to recapitulate the elasticity of all-DNA networks~\cite{xing2019structural}. We incorporate here the building blocks' internal flexibility to enable extensions to mixed-valency architectures and systematic exploration of interaction strength, stiffness, number density and composition.

To accelerate exploration of the large design space and reduce the computational cost of exhaustive parameter scans, we combine simulation outputs with machine-learning surrogate models, specifically Gaussian Process Regression (GPR), capable of interpolating and predicting rheological observables across non-sampled regions~\cite{gadzekpo2025integrative,kadupitiya2020mlsurrogate,zhang2024mlsoftmatter}. 
By comparing predicted results to experimental data, we demonstrate the capacity of the method to identify the appropriate region of the parameter space, where the model yields rheological curves that qualitatively match the experimental ground truth.

Overall, our goal is to provide a scalable and transferable modeling workflow that links motif-level geometry and interaction rules to emergent bulk structure and mechanics, thereby facilitating rational design of attractive soft-matter networks with tailored functionality. 

\section{Model and methods}
\label{sec:methods}

We discuss here in some detail the three elements that comprise our synergistic approach. To give more context, we also introduce the experimental system under investigation, serving as a case study. We consider a binary mixture of DNA nanomotifs, comprised of Y-shaped nanostars and linear linker oligonucleotides~\cite{gadzekpo2025integrative}. By design, both linker and Y-shaped molecules have dangling ends, made of single-stranded DNA; sequence complementarity allows the hybridization of two dangling ends, commonly referred to as ``sticky'' ends, between a linker and a Y-shape only. Following the terminology used in patchy systems, the Y-shapes have a valency of 3 and the linkers have a valency of 2; this design enables selective Y-linker binding, leading to the formation of a network with specific architecture. Further, the valency asymmetry defines a geometric stoichiometric optimum, \textit{i.e.}, a Y:L (Y-shape-to-linker ratio) of $2:3$ ($\mathrm{ratio}=0.667$), ensuring maximum complementary pairing~\cite{Xingetal2010}.

\subsection{Coarse-grained model}

We employ a minimal coarse-grained bead-spring framework to investigate the assembly, melting and viscoelastic response of mixed-valency nanostar-linker networks. 
Following Ref.~\cite{xing2018microrheology}, each unit is represented as a set of ``structural'' and ``patchy'' beads, where the latter model the DNA sticky ends. The units of length and energy of the model are $\sigma$, the excluded volume size of a bead, and the thermal energy $k_B T$, respectively. Y-shaped units are composed by a central bead connected to three arms; each arm is formed by $Y_\mathrm{arm}$ beads ($Y_\mathrm{arm}$ = 2 or 3 in this study) and terminates with a single patch. Linkers are instead modeled as linear chains of 3 beads with patches at both termini (\textit{cf.}, Figure~\ref{fig:1}A).
Intra-molecular connections, involving structural and patch beads that define the building-block geometry, are modeled as harmonic interactions:
\begin{equation}
\label{eq:harm}
V_{\mathrm{bond}}(r) = K_{\mathrm{bond}}(r-r_0)^2,
\end{equation}
where $r$ is the distance between two bonded monomers, $r_0=0.96\sigma$ is the equilibrium bond length and $\beta K_{\mathrm{bond}}=1000$, with $\beta = 1/(k_B T)$, is the bond stiffness~\cite{xing2019structural}. 
Because of excluded volume interactions, non-bonded structural beads repel each other at short range via a truncated and shifted Lennard-Jones (Weeks-Chandler-Andersen, WCA) potential:
\begin{equation}
\label{eq:wca}
V_{\mathrm{WCA}}(r)=
\begin{cases}
4 \epsilon_{\mathrm{LJ}}\left[\left(\dfrac{\sigma}{r}\right)^{12}-\left(\dfrac{\sigma}{r}\right)^{6}\right]+ \epsilon_{\mathrm{LJ}}, & r\le 2^{1/6}\sigma,\\[6pt]
0, & r>2^{1/6}\sigma,
\end{cases}
\end{equation}
where $r$ is distance between beads and $\beta \epsilon_{\mathrm{LJ}}=1$ sets the repulsive energy scale.
Since the length of both the Y-shapes' arms and the linker is small, compared to the persistence length of double-stranded DNA, we employ a bending potential between any triplet of bonded structural beads:
\begin{equation}
 V_{\mathrm{bend}}(\theta) = K_{\mathrm{bend}}(\theta-\theta^{\mathrm{b}}_0)^2,
\end{equation}
where $\theta$ is the angle between the three beads and $\beta K_{\mathrm{bend}}=1000$, ensuring the arms' and linkers' rigidity. The last parameter, $\theta_0$, is the preferred angle, set by the architecture: for triplets within arms or linkers $\theta_0=\pi$, while for triplets that include the central bead in a Y-shape $\theta_0=2\pi/3$~\cite{Liu2024ReactiveBeadSpring}. Finally, the planarity of Y-shapes is enforced via a dihedral-like potential:
\begin{equation}
 V_{\mathrm{planar}}(\chi)= K_{\mathrm{planar}}(\chi-\chi_0)^2,
\end{equation}
where $\chi$ is the angle formed by quadruplets of structural beads that include the central bead in a Y-shape (usually called an ``improper'' angle), $\chi_0=0$ is the preferred angle and $\beta K_{\mathrm{planar}} = 10$ sets the stiffness of the potential.\\
To match the reference DNA system, we distinguish two types of patchy beads based on their assignment to a nanomotif. Patchy beads of the same kind experience a hard-core repulsion; patches and structural beads are connected via harmonic springs. However, patchy beads of different kind used to connect Y-shapes with linkers, also referred to as ``complementary'' patches, interact via a short-ranged attractive potential:
\begin{equation}
\label{eq:att}
 V_{\mathrm{patch}}(r)=
\begin{cases}
4  \varepsilon\left[\left(\dfrac{\sigma_p}{r}\right)^{12}-\left(\dfrac{\sigma_p}{r}\right)^{6}\right]-\beta V_{\mathrm{shift}}, & r\le r_c,\\[6pt]
0, & r>r_c,
\end{cases}
\end{equation}
where $\beta \varepsilon$ is the effective hybridization strength that we vary in our systematic approach, $\sigma_p=0.4\sigma$ sets the interaction range, and $r_c=1.5\sigma$ is the attraction cut-off; the addition of the constant $\beta V_{\mathrm{shift}}$ ensures the continuity of the potential at $r_c$. The attractive patch-patch interaction represents reversible physical self-assembly between complementary binding sites, set by $\beta \varepsilon$; thus, mimicking the reference DNA system, bonds can dynamically form and dissociate due to thermal fluctuations governed by $\beta \varepsilon$ enabling continuous restructuring of the network during assembly.
Finally, we regulate patch flexibility (both on nanomotifs termini) via a separate harmonic bending potential:
\begin{equation}
 V_{\mathrm{angle}}(\theta_{\mathrm{p}})= K_{\mathrm{angle}}(\theta_{\mathrm{p}}-\theta^{\mathrm{p}}_0)^2,
\end{equation}
where now $\theta_{\mathrm{p}}$ is the instantaneous angle between the only triplet of bonded beads that includes the patch (\textit{cf.}, Figure~\ref{fig:1}A),
$\theta_0=\pi$ is the preferred orientation based on energy minimization and $\beta K_{\mathrm{angle}}$ controls orientational flexibility of the patch, also a subject of systematic exploration in our study. Lower values of $\beta K_{\mathrm{angle}}$ allow higher angular freedom, increasing configurational entropy and modulating the effective self-assembly between complementary sites.

\subsection{Simulation details}

We simulate the model using the simulation code LAMMPS~\cite{plimpton1995fast,thompson2022lammps}. As reported in the model description, we set $k_B T=1$ as the unit of energy and $\sigma=1$ as the unit of length; further, we set the mass of a single bead as the unit of mass ($m=1$), which yields the unit of time $\tau=\sqrt{\beta m\sigma^2} = 1$. We perform Langevin Dynamics simulations, integrating the equations of motion using the standard Velocity Verlet algorithm with elementary time step $\Delta t=10^{-3}\tau$. To suppress inertial effects and ensure the overdamped limit, we choose a large value for the friction coefficient $\gamma=100 \tau^{-1}$. We quantify the passage of simulation time in terms of the Brownian time $\tau_B$ defined as the time required for each bead to diffuse its own diameter. Moreover, we simulate systems of different sizes, comprising a total number of particles $N_{p}$ = 20-1500, and we fix the number density, defined as the number of nanomotifs per unit volume, to $\rho = 0.0056 \sigma^{-3}$, corresponding to an approximate experimental DNA concentration of 630 $\mu M$.\\
Initial configurations are generated by placing particles on a uniform grid to avoid overlaps. For each set of input parameters, we first perform short runs in the absence of attractive interactions, allowing spatial and orientational randomization. Afterwards, patch attraction is turned on: we perform sufficiently long simulations (3 $\times$ $10^4 \tau$) to initiate assembly and observe network formation (Supplementary Figure 1); we record the configuration of the system every $1000 \tau$. Due to finite simulation time and diffusion-limited encounters, full connectivity is not always achieved, although it is expected in the long-time limit for sufficiently high values of $\beta \varepsilon$. In practice, we term the system "assembled" when the degree of association, defined below, reaches a plateau. Due to finite simulation time and diffusion-limited encounters, complete bonding corresponding to 100\% connectivity is not always ensured, as complementary sites may remain spatially separated beyond the attractive interaction range. To this end, we adopt a relaxed criterion corresponding to near-complete connectivity ($\approx 98\%$), reflecting an almost fully assembled state obtainable within the probed simulation time. This criterion is exemplified in Supplementary Figure~2, where only very few isolated nanomotifs persist at long times. This leads to the establishment of a quasi-steady state. Extending from this treatment of our system, we then perform long runs (5 $\times$ $10^5 \tau$) to extract steady-state properties, while time-resolved structural metrics are recorded throughout the assembly process.

\subsection{Structural and rheological observables}

We analyze the assembly phase and the steady state of the model by probing both conformational and dynamic quantities.
We compute the time-resolved radial distribution function $g(r)$, defined as:
\begin{equation}
g(r,t)=\frac{1}{4\pi r^2 \rho N_{p}}
\left\langle \sum_{i\neq j}\delta(r-|\mathbf{r}_i(t)-\mathbf{r}_j(t)|)\right\rangle,
\end{equation}
where $r_i(t)$ is the position of the centre of mass of a construct (nanomotif). The time-resolved $g(r,t)$ allows to keep track of the assembly, signaling the emergence of local structural correlations, with short-range peaks reflecting increasing bond formation and the development of connected network environments.\\
We further compute the degree of association (DOA) as a metric for network connectivity. We consider a pair of complementary patchy beads as bonded if their separation distance satisfies the condition $r \le r_{\mathrm{bond}}=\sigma$. Denoting with $M(t)$ the instantaneous number of complementary bonded pairs, we define DOA as:
$\alpha(t) = \frac{M(t)}{M_{\max}}$,
where $M_{\max}=\min\!\left(N_Y f_Y,\; N_L f_L\right)$ is the maximum number of Y-linker bonds given $N_Y$ Y-shapes of valency 3 and $N_L$ linkers of valency 2. By construction, $0\le \alpha \le 1$.
For a given set of input parameters, $\alpha(t)$ will reach a plateau at sufficiently long times, which we take as the condition for the completion of the assembly. The value $\alpha \equiv \alpha(t \to \infty)$ depends on the specific values of the input parameters, \textit{e.g.}, $\beta \varepsilon$ or $\beta K_{\mathrm{angle}}$. In this limit, we define the effective association constant as: 
\begin{equation}
K_a^{*} = \frac{\alpha}{(1-\alpha)^2},
\end{equation}
from which we compute an effective binding free energy:
\begin{equation}
\beta \Delta G_{\mathrm{eff}} = - \ln K_a^{*}
\end{equation}
that provides a compact thermodynamic metric for comparing melting and self-assembling behavior upon variations of the input parameters~\cite{Xia2023Vitrimers}.

Of highest importance for our synergistic approach is the computation of rheological curves. We compute the linear viscoelastic response from equilibrium simulations using the Green-Kubo formalism, based on the shear stress autocorrelation function (SACF). Statistical noise is reduced by averaging over independent off-diagonal components and block-averaging over trajectory segments. We record the off-diagonal components of the microscopic stress tensor, $\Sigma_{\alpha \beta}(t)$ where $\alpha,\beta = x,y,z$ and compute:
\begin{equation}
C_{\alpha\beta}(t)=\left\langle \Sigma_{\alpha\beta}(t_0)\,\Sigma_{\alpha\beta}(t_0+t)\right\rangle,
\label{eq:sacf}
\end{equation}
where $\langle\cdot\rangle$ denotes the lag time average at steady state.
From the SACF we obtain the shear relaxation modulus:
\begin{equation}
G(t)= {\beta V}\,C_{\alpha\beta}(t),
\label{eq:Gt}
\end{equation}
where $V$ is the volume of the simulation box. 
The storage and loss moduli $G'(\omega)$ and $G''(\omega)$ are computed as Fourier sine and cosine transforms of $G(t)$:
\begin{equation}
    G'(\omega) = \omega \int_{0}^{\infty} \sin(\omega t) G(t) dt,  
    \label{eq:Gomega1}
\end{equation}
\begin{equation}
    G''(\omega) = \omega \int_{0}^{\infty} \cos(\omega t) G(t) dt,
    \label{eq:Gomega2}
\end{equation}
where the integrals are evaluated numerically using the discretely sampled $G(t)$. Because long-lag correlations are noisy, we assess convergence by varying the integration window and confirming that the inferred spectra are insensitive to modest changes in the cutoff time. 

We build the relaxation spectrum around the cross-over frequency $\omega_c$, defined by the value of the angular frequency at which $G'(\omega_c)=G''(\omega_c)$, a relation that provides a proxy towards the dominant structural relaxation timescale: lower values of $\omega_c$ correspond to longer relaxation times and a more solid-like response on the timescale of the acquisition.

\subsection{Machine-learning surrogate modeling: GPR and active learning}

To extend rheological predictions beyond the discrete set of simulated state points, we employ GPR as a probabilistic surrogate model, predicting rheological observables as a function of control parameters. Inputs to the GPR model are simulation control parameters (\textit{e.g.}, attraction strength $\beta \varepsilon$, patch stiffness $\beta K_{\mathrm{angle}}$, $\rho$, Y:L, Y-arm length ($Y_\mathrm{arm}$) and system size $N_{p}$); the target is the cross-over frequency $\omega_c$ derived from the Green-Kubo analysis outlined above. GPR is a suitable model choice for this task due to its ability to provide both a mean prediction and an uncertainty estimate, as well as its robust performance in the low-data regime.

For each candidate state point $\mathbf{x}$, GPR provides a posterior distribution $\omega_c(\mathbf{x}) \mid \mathcal{D} \sim \mathcal{N}(\mu(\mathbf{x}), STD^2(\mathbf{x}))$, where $\mu(\mathbf{x})$ is the mean prediction and $STD(\mathbf{x})$ is the posterior standard deviation. We implement uncertainty-based active learning: starting from 4 initial random samples, the algorithm iteratively trains GPR based on new simulation input, computes  $STD(\mathbf{x})$ for remaining candidates and selects the simulation with lowest predictive confidence, which maximizes information gain. 

\subsection{Experimental method}
Experimental characterization of DNA-based hydrogels was performed following established protocols for DNA nanomotif assembly and rheological analysis \cite{xing2018microrheology,stoev2020on,can2025mechanically}. DNA nanomotifs were prepared by thermally annealing stoichiometric mixtures of complementary single-stranded DNA (ssDNA) oligomers in phosphate buffer saline solution (10 mM PB with 100 mM NaCl) of controlled ionic strength. The system was gradually heated above the melting temperature and then cooled to facilitate steady hybridization and minimize the possibility of forming secondary structures. The well-defined multi-valent motifs with complementary sticky ends then self-assemble into network structures, which we refer to as DNA hydrogels, via reversible hybridization dependent on temperature, concentration and sequence design \cite{can2025mechanically}.

The resulting hydrogels were characterized in terms of their thermodynamic and mechanical properties to enable direct comparison with simulations. Melting behavior was determined from temperature-dependent measurements, where the transition corresponds to partial bond dissociation analogous to the $\alpha$ used in simulations \cite{xing2019structural}. The viscoelastic response was quantified using a stress-controlled bulk rheometer (MCR Physica 501, Anton Paar), yielding the storage modulus $G'(\omega)$ and loss modulus $G''(\omega)$ as a function of angular frequency. The inverse of the cross-over frequency provides a measure of the characteristic relaxation time of the network. These experimentally accessible observables establish a direct link to simulation parameters, allowing validation of predicted structure-property relationships across distinct thermodynamic and rheological regimes.

\section{Results}
\label{sec:res}

\subsection{Structural organization and evolution of the Y-linker network}
\label{sec:mod1}

\begin{figure}[]
    \centering
    \includegraphics[width=\textwidth]{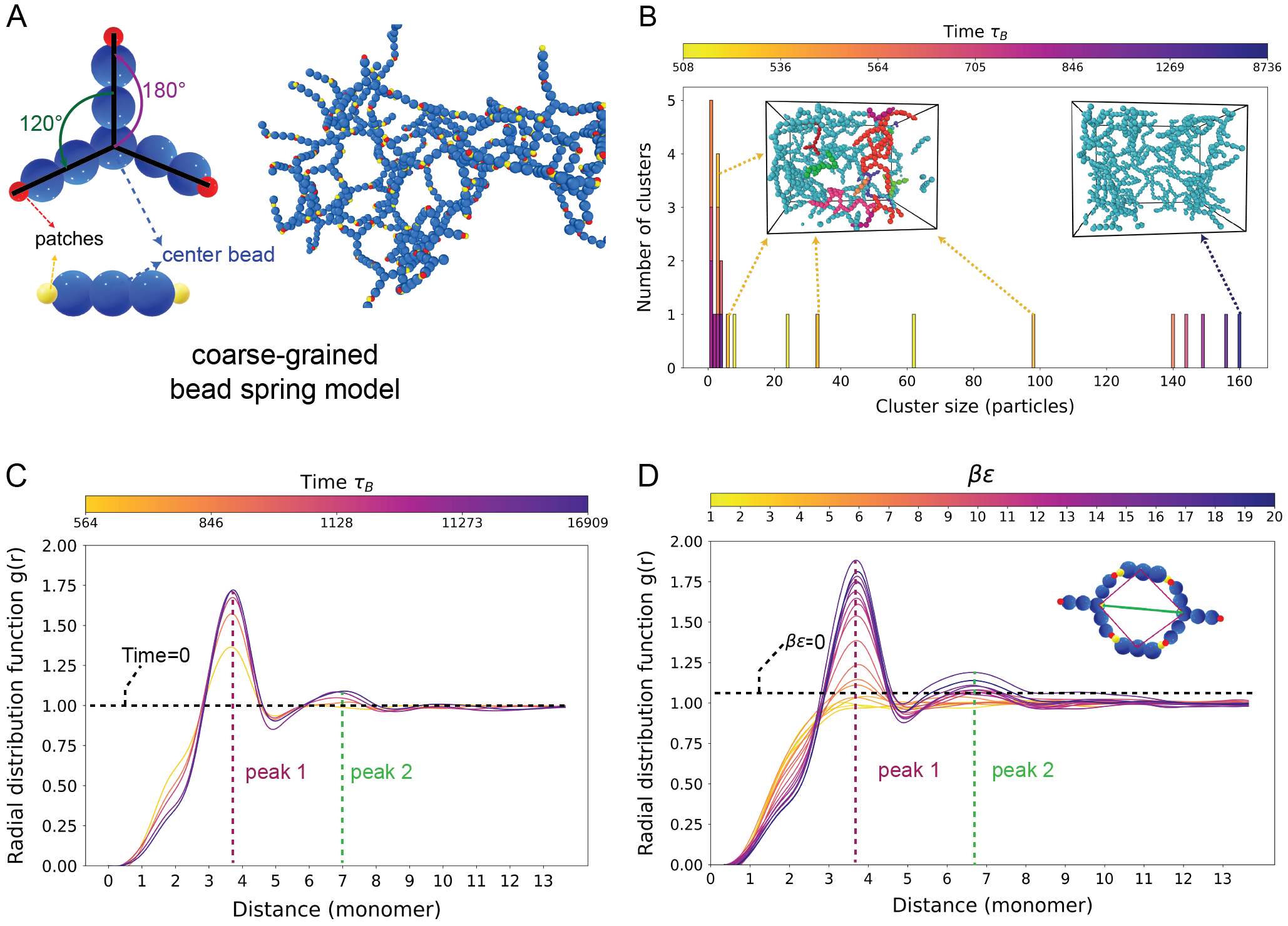}
    \caption{Coarse-grained modeling, assembly kinetics and structural characterization of Y-linker networks.
    (A) Structures of isolated Y-shaped nanostar and linear linker in the bead-spring model (left), accompanied by a snapshot of an assembled network (right).
    (B) Time evolution of the cluster size distribution during assembly at fixed $\beta \varepsilon = 20$. Time is reported in units of the Brownian time $\tau_B = \sigma^2 / D$, where $\sigma$ denotes the bead diameter and $D$ marks the diffusion coefficient. Insets show snapshots of our simulation at intermediate and late stages.
    (C) Time evolution of the radial distribution function $g(r)$ at fixed $\beta \varepsilon = 20$, showing the emergence of local structural correlations as inter-particle bonding progresses; the dashed line marks the ideal gas reference.
    (D) Long time radial distribution functions for several values of $\beta \varepsilon$, demonstrating enhanced short-range correlations and structural ordering at higher attraction strength; the dashed curve labeled $\beta \varepsilon=0$ denotes the uncorrelated baseline (\textit{cf.}, panel C).}
    \label{fig:1}
\end{figure}

We study the assembly process, \textit{i.e.}, the emergence of structure, by tracing how an initially uncorrelated mixture evolves into a connected network. Specifically, we characterize the assembly across multiple lengthscales, employing the structural observables introduced in Section~\ref{sec:methods} (\textit{cf.}, Figure~\ref{fig:1}).
We begin by examining the evolution of the cluster size distribution, capturing how connectivity develops during assembly. Figure~\ref{fig:1}B shows the time-dependent distribution of cluster sizes at fixed $\beta \varepsilon = 20$~\cite{Newman2010}. 
At early times, the system is dominated by monomers and small clusters, reflecting limited bond formation. As time progresses, the distribution broadens, indicating the presence of multiple finite clusters that continuously merge and reorganize. At later times, we observe the emergence of a system-spanning cluster, signaling the onset of percolation (Supplementary Figure 2). This cluster continues to evolve as additional bonds form between available binding sites. In the final stage of the assembly, as defined in Section~\ref{sec:methods}, the system approaches maximum connectivity with a stable bond network: the late-time configuration in Figure~\ref{fig:1}A showcases a disordered, space-filling network in which most motifs participate in a single connected component, consistent with gel-like behavior in patchy systems.

The inspection of the development of local structural correlations during this process yields further insight. Figure~\ref{fig:1}C shows the time evolution of the radial distribution function $g(r)$, which tracks the emergence of local order as bonding progresses. At $t=0$, the system shows negligible local structure. As directional interactions drive assembly, a pronounced first peak emerges, reflecting nearest-neighbor correlations self-assembled with bonded Y-linker contacts. A weaker second feature at larger distances indicates the development of correlations beyond the first coordination shell, consistent with the formation of locally connected nanomotifs. The growth and sharpening of these features provide a direct structural signature of increasing local organization during assembly. These results show that local correlations arise early in the assembly process and precede the formation of a fully connected network, consistent with expected behavior in self-assembled systems and indicating that the model reproduces the anticipated microscopic assembly pathway.\\
Following previous studies~\cite{StaufferAharony1994, rovigatti2018simulate,xing2019structural}, we highlight the role of interaction strength $\beta \varepsilon$ in determining the structure of the assembled network (Figure~\ref{fig:1}D). In the absence of patch-patch attraction ($\beta \varepsilon = 0$), the radial distribution function remains nearly flat, as expected for a quasi-hard-sphere fluid at low density.
As $\beta \varepsilon$ increases, the first peak becomes significantly higher and sharper, reflecting an increased probability of bonded neighbors. As before, the enhancement of the second peak further indicates the development of extended structural correlations, as a stronger patch-patch attractive interaction promotes more stable and interconnected local environments: increasing $\beta \varepsilon$ not only stabilizes bonds but also enhances the degree of local structural organization, leading to a more robust and densely connected network.

\subsection{Thermodynamics of self-assembly and melting}
\label{sec:mod2}

\begin{figure}[]
    \centering
    \includegraphics[width=\textwidth]{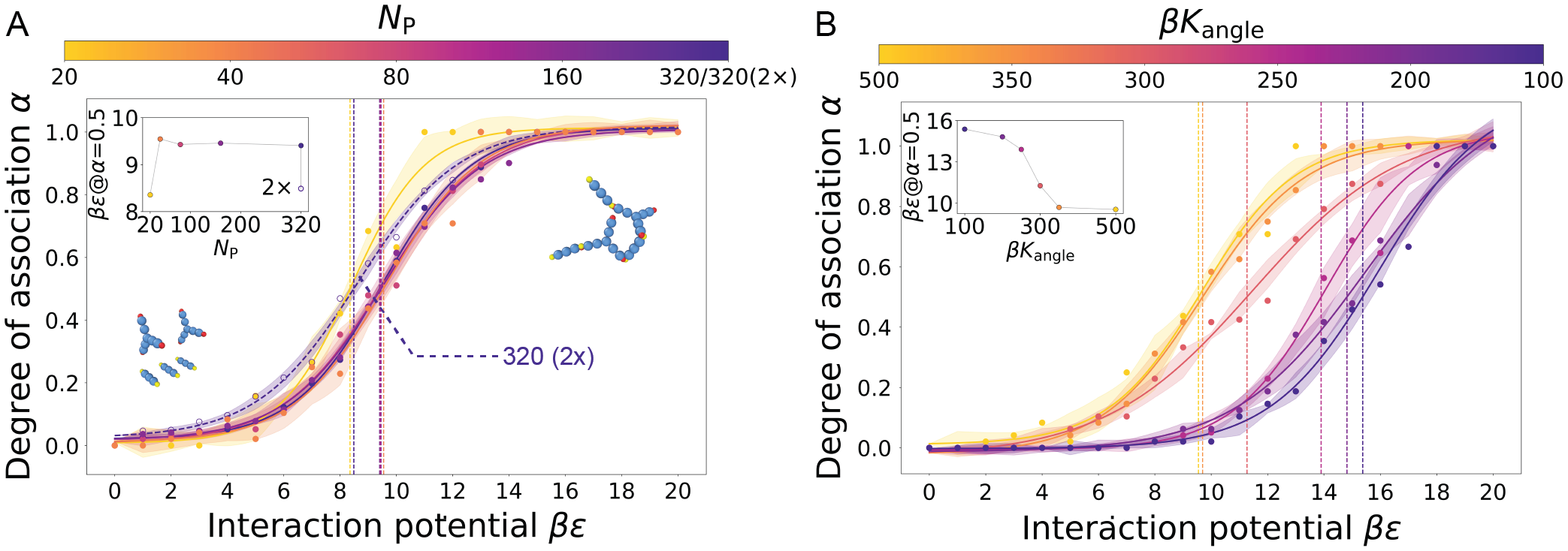}
    \caption{
    Thermodynamics of self-assembly in rigid and flexible systems.
    Long-time DOA $\alpha$ as a function of interaction strength $\beta \varepsilon$.
    (A) System with no patch flexibility, undergoing melting at different system sizes and $\rho$. The effective melting point coincides with $\mathrm{\alpha} = 0.5$ (vertical dashed lines). The behavior converges for systems of >40 nanomotifs (inset). Increasing the $\rho$ shifts the melting point to lower $\beta\varepsilon$, corresponding to a higher melting temperature.
    (B) System with flexible sticky patches at a fixed number of particles. Patch flexibility is controlled by the stiffness parameter $\beta K_{\mathrm{angle}}$. Increasing flexibility (lower $\beta K_{\mathrm{angle}}$) shifts the melting transition to higher $\beta\varepsilon$ (inset), indicating that a higher value of the dimensionless interaction strength is required for self-assembly. This trend is consistent with experimental observations for DNA systems with flexible sticky ends~\cite{stoev2020on}.
    }
    \label{fig:Tm:bulk_vs_flexible}
\end{figure}

We characterize the thermodynamics of self-assembly employing the long-time $\alpha$, defined in Section~\ref{sec:methods}, that serves as an order parameter to distinguish between a dilute, weakly interacting fluid and an assembled, connected network~\cite{StaufferAharony1994}.

Figure \ref{fig:Tm:bulk_vs_flexible}A shows $\alpha$ for a system with low patch flexibility, as defined in Section~\ref{sec:methods} and specifically for $\beta K_{\mathrm{angle}}=500$, as a function of $\beta \varepsilon$ for systems of different sizes, \textit{i.e.}, different number of particles $N_p$ at a fixed density.
All the $\alpha$-$\beta \varepsilon$ curves in Figure~\ref{fig:Tm:bulk_vs_flexible}A exhibit a sigmoidal shape, closely resembling melting curves observed in DNA hybridization experiments~\cite{xing2018microrheology,stoev2018using,can2025mechanically}.
At low enough values of $\beta \varepsilon$, thermal fluctuations dominate over patch-patch attraction: particles remain mostly unbound, resulting in low values of $\alpha$. As $\beta \varepsilon$ increases, bonds between complementary patches become increasingly long-lived, leading to a rapid rise in $\alpha$ and the formation of an extended network.
With increasing $N_p$, the melting curves converge on the same master curve, indicating the emergence of bulk-like behavior;
for small systems ($N_p \leq 40$), the melting curves display noticeable finite-size effects. 
Following the analogy with DNA hybridization experiments, we define an effective melting point as the value of the interaction strength at which ${\alpha} = 0.5$, corresponding to the state where on average half of all possible bonds are formed or broken. 

An interesting feature of the chosen model is the possibility to tune patch flexibility through the parameter $\beta K_{\mathrm{angle}}$: we investigate its role on the melting curves. In DNA-based systems, patch flexibility can be introduced through the presence of free bases in single-stranded DNA sticky ends~\cite{can2025mechanically}. The melting curves are shown in Figure~\ref{fig:Tm:bulk_vs_flexible}B.
As $\beta K_{\mathrm{angle}}$ is reduced, the melting transition shifts systematically to higher values of $\beta \varepsilon$, indicating that stronger attraction or, equivalently, lower temperature is required to achieve the same $\alpha$. We attribute this behavior to the increased configurational entropy of the unbound state in the case of high patch flexibility. Specifically, forming a bond requires a significant reduction in the local number of available configurations of the patch and, when patches are highly flexible, the self-assembled entropic penalty is greater. Consequently, a larger value of $\beta \varepsilon$ is necessary to compensate this loss of entropy and stabilize the bonded state. 

Interestingly, this behavior closely resembles experimental findings for {DNA-based hydrogels}, where increasing the number of unpaired {nucleotide spacers} adjacent to the sticky ends lowers the melting temperature~\cite{stoev2020on,can2025mechanically}. Overall, these results confirm that the model captures key thermodynamic trends observed in experiments, including the effect of sticky-end flexibility on the melting behavior.

\subsection{\textit{In-silico} rheology}
\label{sec:mod3}

\begin{figure}[]
    \centering
    \includegraphics[width=\textwidth]{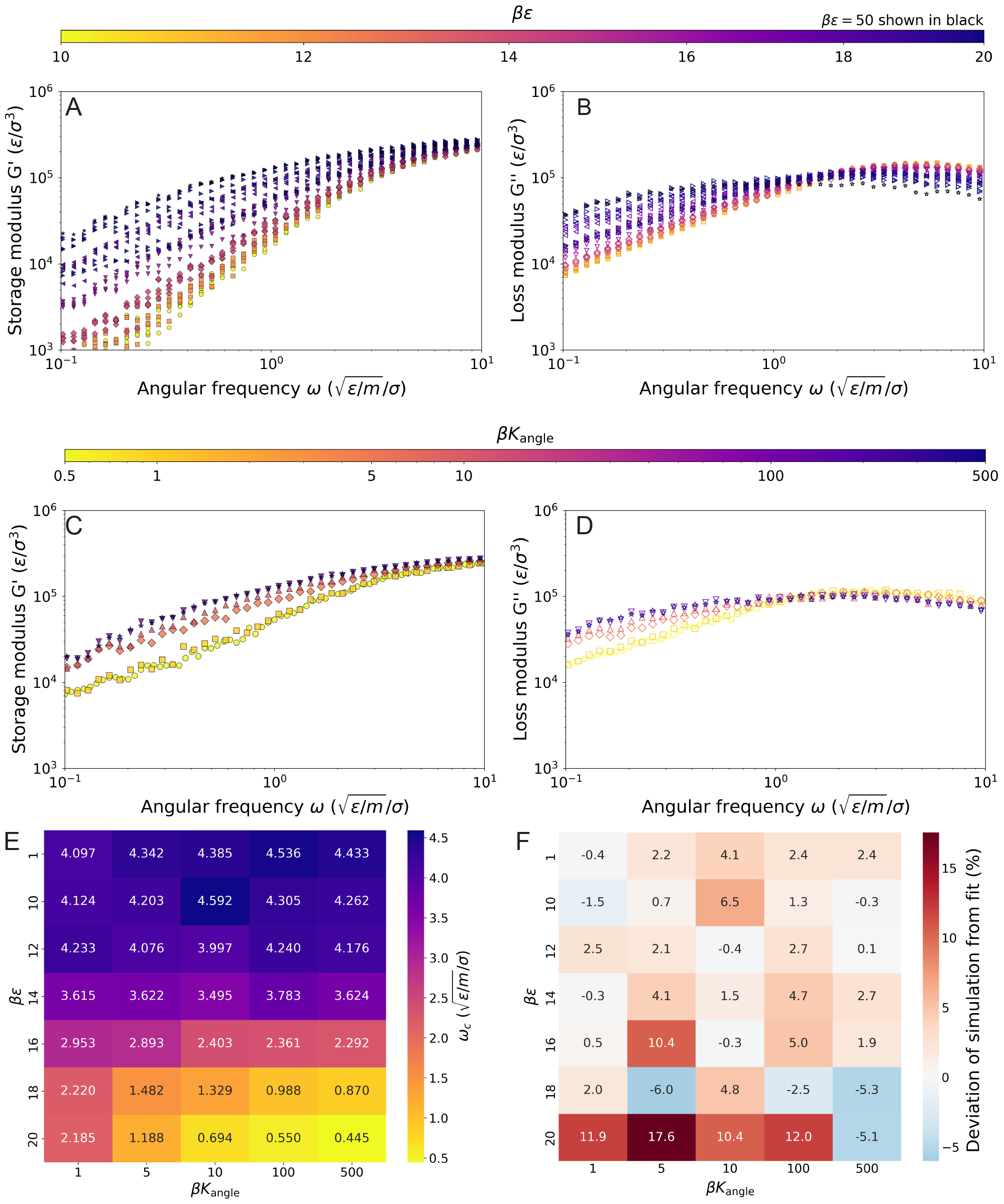}
    \caption{Parameter dependence of rheological properties across $(\beta \varepsilon, \beta K_{\mathrm{angle}})$ space.
    (A-B) Effect of interaction strength $\beta \varepsilon$ on $G'$ and $G''$ (fixed $\beta K_{\mathrm{angle}}$ = 500, Y:L ratio = 0.667, $Y_{\mathrm{arm}}$ = 2, $\rho$ = 0.00333, $N_p$ = 1,280).
    (C-D) Effect of patch flexibility $\beta K_{\mathrm{angle}}$ on $G'$ and $G''$ at fixed $\beta \varepsilon=20$, other parameters as in A-B.
    (E) Heatmap of cross-over frequency $\omega_c$ across the full $(\beta \varepsilon, \beta K_{\mathrm{angle}})$ parameter space showing regime transition at $\beta \varepsilon = 14$.
   (F) Mean relative deviation between the cross-overs obtained in panel E from raw simulation data and those obtained from fitted rheological curves (2.70\%), quantifying simulation data quality and level of random noise in the conversion from particle dynamics.}
    \label{fig:param_sweep}
\end{figure}

We now turn to the dynamic response of the assembled networks. To this aim, we compute the SACF as  described before in Section~\ref{sec:methods}, using the Green-Kubo formalism~\cite{Green1954, Kubo1957}. The SACF (\textit{cf.}, Supplementary Figure 3A) exhibits two clearly separated regimes: a fast exponential decay arises at short times from local particle rearrangements, while a much slower decay emerges at longer times from collective network relaxation. This slow component controls the viscoelastic behavior and marks a transition between a viscous-dominated and an elastic-dominated response~\cite{Murashima2021ViscosityOvershoot}.

From the SACF we obtain the time-dependent complex shear modulus, which we decompose into real (storage, $G'$) and imaginary (loss, $G''$) parts (Eq.~\eqref{eq:Gomega1} and ~\eqref{eq:Gomega2}, \textit{cf.}, Supplementary Figure 3B). The cross-over frequency $\omega_c$ provides a scalar measure of the structural relaxation time and marks the transition from fluid-like to solid-like behavior~\cite{FitzSimons2020ForwardKineticsHydrogels,Xia2023Vitrimers}. 

Figure~\ref{fig:param_sweep} presents a comprehensive analysis of how microscopic parameters control macroscopic rheological behavior.
Panels A-B show the effect of interaction strength $\beta \varepsilon$ on storage and loss moduli (at fixed $\beta K_{\mathrm{angle}}=500$): as $\beta \varepsilon$ increases, both moduli shift toward more elastic behavior, with $G'$ dominating at lower frequencies. On the other hand, panels C-D illustrate the effect of patch constraint parameter $\beta K_{\mathrm{angle}}$ on the moduli (fixed $\beta \varepsilon=20$): stiffer bonds enhance elastic response primarily at high frequencies. Comparing the dependence of the viscoelastic moduli on $\beta \varepsilon$ and $\beta K_{\mathrm{angle}}$ we find an overall higher order scaling with interaction strength: therefore, with regards to tuning the mechanical behavior $\beta \varepsilon$ and $\beta K_{\mathrm{angle}}$ can be perceived as a ``coarse knob'' and a ``fine knob'', allowing careful adjustment to experimental parameters.

Panel E presents a heatmap of $\omega_c$ across the full $(\beta \varepsilon, \beta K_{\mathrm{angle}})$ parameter space, revealing two distinct regimes separated by $\beta \varepsilon \approx 14$. Below this threshold, $\omega_c$ shows weak parameter dependence, corresponding to semi-associated networks with short bond lifetimes. Above $\beta \varepsilon = 14$, the system becomes fully associated, consistent with the structural signatures observed in Figure~\ref{fig:1}, where the formation of a system-spanning network and the development of pronounced short-range correlations reflect near-complete bonding at the microscopic level. Generally, $\omega_c$ is found to be highly sensitive to both $\beta \varepsilon$ and $\beta K_{\mathrm{angle}}$, decreasing significantly as interactions strengthen and bonds stiffen. Our discrete simulation sampling does not allow a high-fidelity estimate of the cross-over frequency. To this end, we resorted to using least-squared fits that interpolate between simulated data points in a way that minimizes the deviations in a plot of residuals. Panel F quantifies the average deviation of our simulations with respect to those fits at different $\beta \varepsilon$ and $\beta K_{\mathrm{angle}}$. On average, we find a deviation of 2.70\% across all simulated conditions (\textit{cf.}, Supplementary Figure 4).

Moreover, we performed a comprehensive one-at-a-time (OAT) sensitivity analysis quantifying the isolated effect of each parameter on $\omega_c$. By varying one parameter at a time while fixing others at reference values ($\beta \varepsilon=20$, $\beta K_{\mathrm{angle}}=500$, Y:L = 0.667, $Y_{\mathrm{arm}}$ = 2, $\rho$ = 0.00333, $N=1280$, yielding $\omega_{c,\text{ref}}=0.469$), we established a parameter hierarchy in terms of influence on $\omega_{c}$ (\textit{cf.}, Supplementary Figure 5): \textbf{(1)} Interaction strength $\beta \varepsilon$ dominates with as high as 8-fold maximum variation ($\beta \varepsilon=10 \to 20$); \textbf{(2)} patch constraint parameter $ \beta K_{\mathrm{angle}}$ exhibits >3-fold variation; \textbf{(3)} Y:L shows nearly 2-fold variation with optimum at the stoichiometric value of 2:3 (ratio = 0.667) corresponding to nearly complete complementary pairing, whereas deviations from stoichiometry leave either excess free ends or frustrated binding sites, lowering network formation efficiency; \textbf{(4)} Y-arm length ($Y_{\mathrm{arm}}$) displays negative correlation with approximately two-fold decrease (2$\to$3 beads per arm) due to topology-driven stiffening, independent of interaction energy; \textbf{(5-6)} $\rho$ and system size contribute $<20\%$, where the latter validates the bulk convergence obtained earlier in Figure~\ref{fig:Tm:bulk_vs_flexible}A.

\subsection{Machine learning for efficient parameter space exploration}
\label{sec:ml1}

Despite the highly coarse-grained nature of the simulations, the system is still characterized with quite a few degrees of freedom. The parameter space we explore spans $\beta \varepsilon \in [10,20]$, $\beta K_{\mathrm{angle}} \in [1,1000]$, multiple number densities $\rho$, Y:L ratios, system sizes and Y-arm lengths, yielding a large parameter space (\textit{cf.}, Supplementary Table 1). With each simulation requiring 672 core hours and the large degree of combinations, exhaustive exploration is computationally prohibitive. To address this challenge, we employ GPR (\textit{cf.}, Section~\ref{sec:methods}) with uncertainty-driven active learning to efficiently navigate the parameter space.

\begin{figure}[H]
    \centering
    \includegraphics[width=\textwidth]{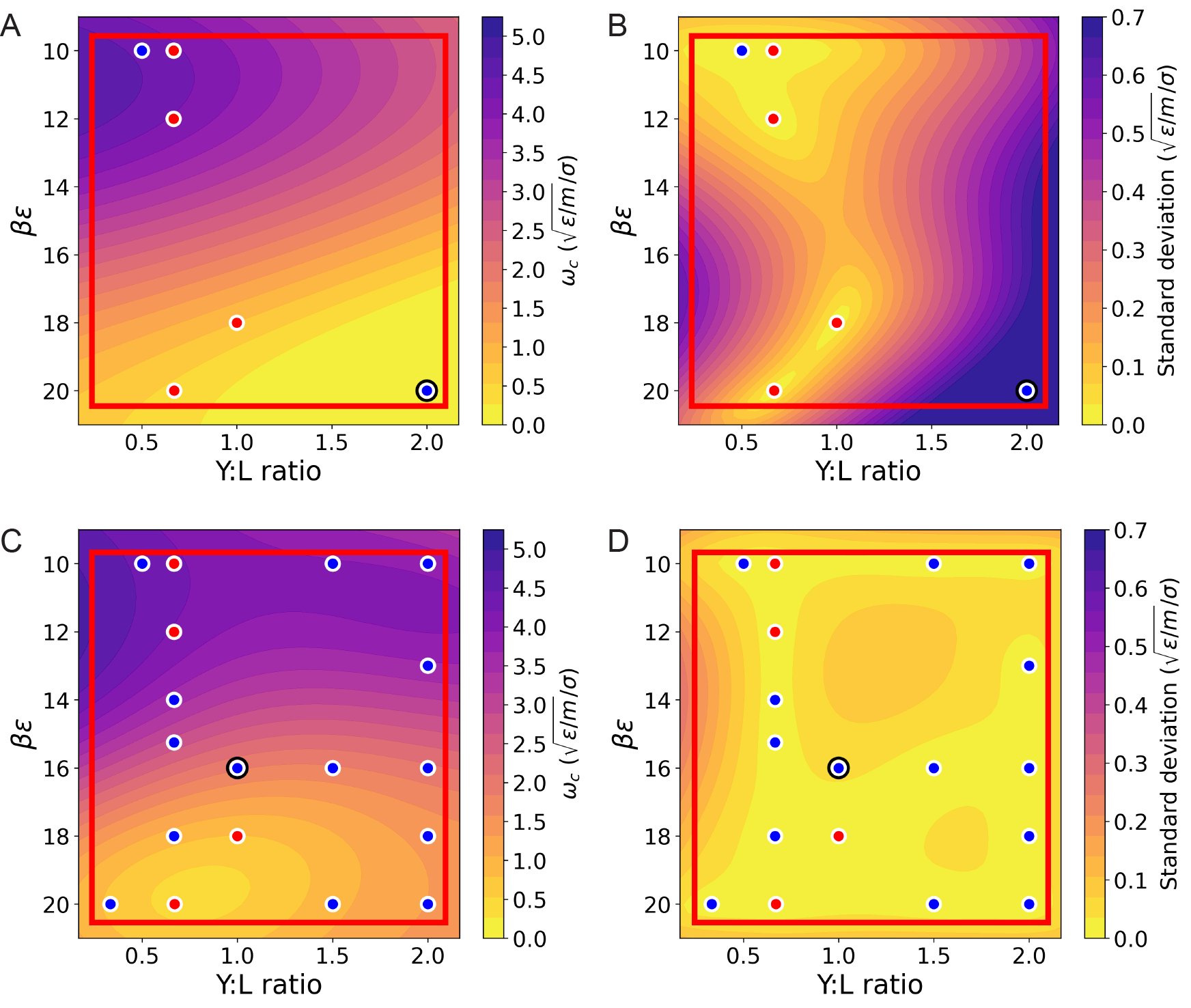}

\caption{Evolution of Gaussian Process active learning in $(\text{Y:L}, \beta \varepsilon)$ space. (A) Predicted $\omega_c$ mean surface at iteration 2 with 6 training samples (4 initial in red, 2 from active learning in blue). (B) Posterior uncertainty at iteration 2 (max $\approx$ 0.96). (C) Predicted $\omega_c$ surface at iteration 14 with 18 training samples (4 initial in red, 14 from active learning in blue), resolving fine-scale structure including stoichiometric minimum at ratio = 0.667. (D) Posterior uncertainty at iteration 14 (mean $\approx$ 0.044, max $\approx$ 0.16). The bold circled point indicates the next simulation to be executed (iteration 14), selected based on maximum uncertainty. Red boxes delineate the region of applicability in which the model interpolates with high confidence; outside this region, the model extrapolates with reduced reliability. The standard deviation quantifying the prediction precision of the machine learning algorithm.}    \label{fig:gpr_ratio_varepsilon}
\end{figure}

Figure~\ref{fig:gpr_ratio_varepsilon} demonstrates the learning trajectory in $(\text{Y:L}, \beta\varepsilon)$ space, the most complex case due to non-monotonic ratio effects. We start with 4 initial random samples and iteratively select the most informative simulations using uncertainty-driven active learning for fixed reference values of secondary parameters ($\beta K_{\mathrm{angle}}$ = 500, $Y_{\mathrm{arm}}$ = 2, $\rho$ = 0.00333, $N_p$ = 1280). Early-stage predictions (panels A-B, iteration 2 with 6 training samples: 4 initial in red + 2 from active learning in blue) capture only coarse trends. The mean surface $\mu$ (panel A) misses fine-scale structure and the uncertainty map (panel B) exhibits large variance regions with maximum standard deviation of approximately 0.96, far from sparse training data. Model performance at this stage is moderate: mean absolute error (MAE) = 0.493, and coefficient of determination (R²) = 0.692 on held-out test set.

At iteration 13 (panels C-D, 18 training samples: 4 initial in red + 14 from active learning in blue), the model achieves high predictive accuracy with MAE = 0.066 and R² = 0.996. The predicted surface (panel C) resolves fine-scale structure including the non-monotonic minimum near the stoichiometric ratio of 0.667, where complete pairing of complementary sticky ends (2:3 Y:L balance) maximizes network connectivity. The uncertainty map (panel D) shows substantially reduced variance, with mean standard deviation of approximately 0.044 (\textit{ca.} 95\% reduction from iteration 2) and maximum standard deviation of approximately 0.16 confined to underexplored corners. The circled point in bold (panel C) indicates the next simulation selected by active learning for iteration 14, demonstrating the algorithm's systematic exploration strategy. Active learning therefore enables efficient exploration of high-dimensional rheological parameter spaces while substantially reducing computational cost.

The non-uniform distribution of training points (panel C) reflects adaptive resource allocation: higher sampling density appears where the model required more information to reduce uncertainty, with sparser coverage in regions exhibiting steadily varying behavior. For uniform grid coverage at resolution $\Delta\varepsilon = 0.25$ and $\Delta\text{ratio} = 0.10$, exploration of the same parameter ranges would require 41 × 18 = 738 simulations. Achieving comparable predictive performance (R² > 0.99) using only 18 strategically selected samples represents 41 $\times$ computational speedup (97.6\% reduction) compared to exhaustive parameter sweeps.

Validation using Gradient Boosting Regression as an alternative ML approach confirms robustness (Supplementary Figure 6). The integrated framework enables rational design of limited-valency soft-matter networks, \textit{e.g.}, DNA hydrogels, with targeted rheological properties using fewer than 20 strategically selected simulations (specifically 18 in this work).

\subsection{Benchmarking \textit{in-silico} rheology against \textit{in-vitro} experiments}
\label{sec:comparison}

To assess whether the coarse-grained model captures experimentally relevant behavior, we compare simulation-derived rheological spectra with experimental data from DNA hydrogel systems YLS6-YLS9, with sequences as described by Can \textit{et al.} ~\cite{can2025mechanically} These systems, similarly comprised of trivalent Y-shapes and bivalent linkers, differ ever so slightly in the length of the sticky ends of the linkers, where in LSX, X denotes the number of bases included in the sticky ends of the linkers. The latter enable hybridization to the sticky ends of the Y-shapes, which in turn dictates the mechanical behavior upon heating and cooling. All hydrogels  exhibit characteristic viscoelastic cross-over measurable via oscillatory rheometry. Shorter linkers tend to form weaker networks characterized by depreciated melting temperatures (sol-gel transition points) and reduced mechanical stiffness under identical experimental conditions.

The comparison strategy employs a semi-automatic mapping: simulations spanning the full $(\beta \varepsilon, \beta K_{\mathrm{angle}})$ parameter space are screened against each experimental system by comparing cross-over frequencies and qualitative curve shapes. Best-match pairs are identified based on minimum deviation criterion, \textit{i.e.}, we minimize the relative error between the cross-over frequency measured in simulations and experiments $\min(\lvert(\omega_{c,sim} - \omega_{c,exp}) / \omega_{c,exp}\rvert)$. To enable qualitative comparison, the simulation cross-over modulus is re-scaled to match the experimental one ($G_{c,\text{sim}}/G_{c,\text{exp}}$), accounting for differences in absolute scale between simulation and experiment, while preserving frequency-dependent trends.

\begin{figure}[H]
    \centering
    \includegraphics[width=\textwidth]{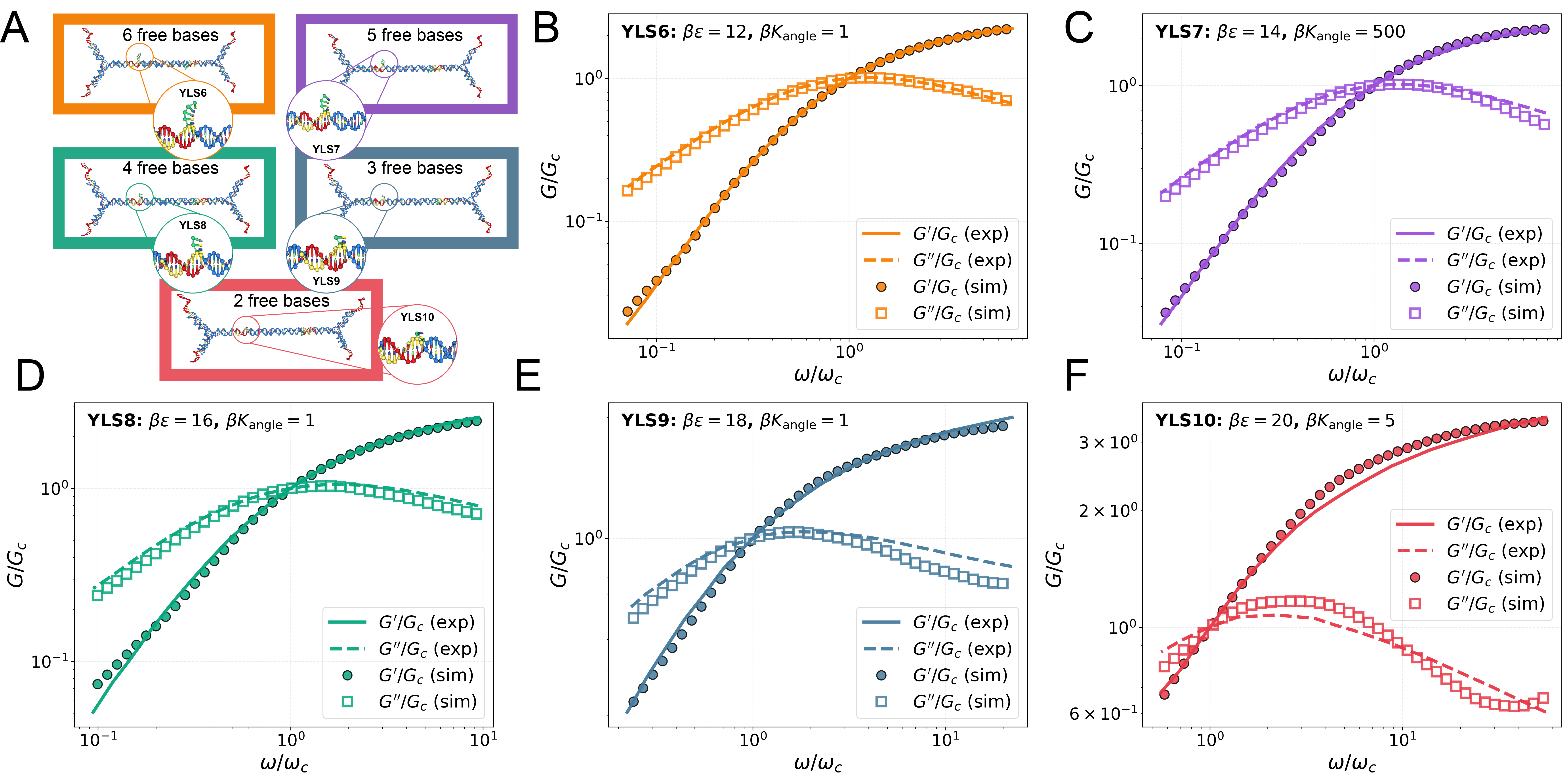}
    \caption{Comparison of experimental and simulated viscoelastic responses for DNA hydrogels with systematically varied linker flexibility. The storage and loss moduli are shown in dimensionless form, normalized by the cross-over coordinates, $\omega/\omega_c$ and $G/G_c$, such that the experimental cross-over is positioned at $(\omega=\omega_c,G=G_c)$. Solid and dashed lines denote the experimental storage modulus, $G'/G_c$, and loss modulus, $G''/G_c$, respectively, while filled circles and open squares show the corresponding simulation results. (A) Schematic overview of the corresponding YLS6-YLS10 linker constructs, containing 6, 5, 4, 3 and 2 free bases, respectively, with enlarged views highlighting the local flexible-joint geometry. Based on a minimum deviation criterion: (B) YLS6 matches with $\beta\varepsilon = 12$ and $\beta K_{\mathrm{angle}} = 1$;  (C) YLS7 matches with $\beta\varepsilon = 14$ and $\beta K_{\mathrm{angle}} = 500$; (D) YLS8 matches with $\beta\varepsilon = 16$ and $\beta K_{\mathrm{angle}} = 1$; (E) YLS9, matches with $\beta\varepsilon = 18$ and $\beta K_{\mathrm{angle}} = 1$;  (F) YLS10 matches with $\beta\varepsilon = 20$ and $\beta K_{\mathrm{angle}} = 5$. Across all systems, the coarse-grained model reproduces the characteristic frequency-dependent shapes of both moduli and captures the shift in viscoelastic response in accordance with the experimental linker design.}
    \label{fig:experimental_validation}
\end{figure}

Figure~\ref{fig:experimental_validation} presents five successful mappings (YLS6-YLS10) to simulations with $\beta \varepsilon \in [12,20]$ and $\beta K_{\mathrm{angle}} \in [1,500]$. The dimensionless representation, with axes normalized by cross-over coordinates ($\omega/\omega_c$,$G/G_c$), enables direct quantitative comparison by collapsing all cross-overs to the universal point $(\omega=\omega_c,G=G_c)$. This representation eliminates scale-dependent variations while preserving the frequency-dependent shape of rheological spectra, facilitating assessment of structural similarities across systems. The successful mapping of these experimental systems has three critical implications. First, it confirms that the coarse-grained model, despite its simplifications, captures the essential physics governing viscoelastic response in these DNA hydrogels: the model correctly describes the shape of rheological spectra in the transition region. Second, the correlation between YLSX and best-match $\beta \varepsilon$ (YLS6$\to$10, $\beta \varepsilon$12$\to$20) suggests that our coarse-grained parameter $\beta\varepsilon$ effectively captures the thermodynamic driving force for DNA base-pairing, with increased sticky-end length requiring proportionally higher interaction energy. Third, the variation in $\beta K_{\mathrm{angle}} \in [1,500]$ reflects differences in effective linker flexibility across experimental systems, where synthetic point mutations (sticky-end deletions) modulate patch constraint across experimental systems. We find it is often required to tune both $\beta \varepsilon$ and $\beta K_{\mathrm{angle}}$ to capture actual experimental changes to the design of the oligos comprising the selected set of hydrogels. This suggests that in establishing future parallels between \textit{in-silico} and \textit{in-vitro} results, one likely needs to resort to multi-parameter mapping, as opposed to 1:1 parameter correspondence.

Further, we found that our probed simulation parameters allow for close matching with five out of seven experimental systems characterised by Can \textit{et al.}~\cite{can2025mechanically} The experimental data for YLS11 and YLS12 show no match within the probed ranges of $\beta \varepsilon$ and $\beta K_{\mathrm{angle}}$, as these systems do not exhibit observable cross-over in the measured frequency range. Withing the simulated ranges of $\beta \varepsilon$ and $\beta K_{\mathrm{angle}}$ we successfully recapitulate the behavior of the DNA hydrogel systems close to melting or the sol-gel transition point. Away from melting and deeper into the solid phase, we expect that the model would require much higher values for the interaction strength and patch rigidity than the ones used here. Additionally, we expect that non-ergodic effects may interfere with recapitulating the experimental viscoelastic properties and more sophisticated level of modeling might be needed to fully capture the locked-in gel-like response. The obtained partial validation sets rough criteria for when coarse-grained predictions are expected to hold and when employing more detailed (and more computationally expensive) all-atom or nucleotide-level oxDNA simulations might be necessary~\cite{Biffi2013,Rovigatti2014,Jamali2015,xing2019structural}.

\subsection{Discussion}

From physical perspective, we find the patch attraction strength $\beta \varepsilon$ to be the primary control knob in our simulations for tuning the mechanical response because it directly sets bond stability and, through bond lifetime, the timescale over which stress can relax. Our results consistently identify two regimes separated by a threshold near $\beta \varepsilon \approx 14$. Below this value the system remains semi-self-associated: connectivity is transient, bond lifetimes are short and the cross-over frequency $\omega_c$ depends only weakly on $\beta \varepsilon$ and the patch stiffness $\beta K_{\mathrm{angle}}$. In this regime, the material response is fluid-like on long timescales, with relaxation dominated by reversible bond breaking and recombination. Above $\beta \varepsilon \approx 14$, the network becomes fully associated, with most complementary patches engaged and a persistent, system-spanning backbone. Here the viscoelastic spectrum develops a pronounced elastic response and $\omega_c$ decreases sharply, indicating substantially slower relaxation. Importantly, once the network is fully connected, the patch constraint parameter $\beta K_{\mathrm{angle}}$ becomes a strong secondary control parameter: increasing $\beta K_{\mathrm{angle}}$ suppresses local bending fluctuations, reduces the number of accessible microstates compatible with bonded configurations and thereby further slows stress relaxation.

The structural diagnostics provide a mechanistic link between these rheological regimes. Time-resolved radial distribution functions show that local correlations emerge early in the assembly process, providing a microscopic predictor for the subsequent assembly of finite clusters and the onset of percolation. Cluster statistics then connect this local ordering to mesoscale coarsening, culminating in the dominance of a single and large connected component limited by the system size. These observations support a consistent picture in which increasing $\beta \varepsilon$ first stabilizes local bonded environments (growth and sharpening of short-range features in $g(r)$) and then drives network formation into a percolated backbone that can sustain stress over long times. Stress autocorrelation analysis confirms that the slow relaxation mode strengthens with increasing $\beta \varepsilon$, consistent with the shift toward solid-dominated response observed in the frequency-dependent moduli.

On the data-driven side, GPR shows that these trends are embedded in a strongly non-linear, high-dimensional structure-property relationship. Active learning achieves substantial efficiency gains: between iteration 2 (6 samples, R² = 0.692) and iteration 14 (18 samples, R² = 0.996), the approach reduces computational cost 41 $\times$ (97.6\% reduction) compared to uniform grid sampling while maintaining predictive precision. One-at-a-time sensitivity analysis (Supplementary Figure 5) complements this by ranking design variables according to their impact on $\omega_c$, with $\beta \varepsilon$ dominating the response across multiple orders of magnitude, followed by $\beta K_{\mathrm{angle}}$ and composition.

Comparison with experimental DNA hydrogel data (Figure~\ref{fig:experimental_validation}) validates that simulation-derived structure-property relationships translate to real materials. The successful mapping of five experimental systems to specific $(\beta \varepsilon, \beta K_{\mathrm{angle}})$ combinations confirms that the coarse-grained framework captures essential viscoelastic physics where mesoscopic interactions dominate near the sol-gel transition point.

Overall, the combined simulation-ML-experiment framework provides a transferable route to rational design of self-assembling soft-matter networks. Structural diagnostics reveal assembly mechanisms, SACF-based rheology translates connectivity into mechanical descriptors, uncertainty-guided learning enables efficient exploration, and experimental validation defines applicability boundaries. Future work can extend this approach by incorporating sequence-dependent interactions and multi-component architectures, enabling closed-loop material discovery where simulations, ML and targeted experiments converge iteratively to strengthen the predictive connection between microscopic design and macroscopic performance.

\section{Conclusion}

We have developed a scalable coarse-grained framework that links microscopic interaction rules to emergent structure and viscoelastic response in mixed-valency Y-linker networks. By systematically varying the patch attraction strength $ \beta \varepsilon$ and patch constraint parameter $\beta K_{\mathrm{angle}}$, we demonstrate that gelation and mechanical response are governed by a coupled effective free-energy balance between energetic stabilization and orientational entropy. The transition from a semi-associated, fluid-like regime to a fully connected, gel-like network does not occur at a single universal $\beta \varepsilon$, but depends sensitively on patch flexibility: stiffer patches (larger $\beta K_{\mathrm{angle}}$) lower the effective association threshold, whereas more flexible patches require stronger attractions to achieve comparable degrees of bonding ($\alpha \approx 0.5$). This highlights the thermodynamic interplay between bond energy and orientational entropy in controlling network formation.

 Structural diagnostics, including the radial distribution function and cluster statistics, reveal how local bonding correlations precede network formation  and ultimately determine macroscopic rheology. The cross-over frequency $\omega_c$ provides a compact descriptor of the resulting relaxation spectrum and clearly distinguishes weakly associated from highly connected states. Once a persistent backbone forms, both $\beta \varepsilon$ and $\beta K_{\mathrm{angle}}$ significantly influence stress relaxation, with stronger attractions and higher stiffness producing slower dynamics and a more pronounced elastic response.

The framework we introduce here is transferable to broader classes of patchy soft-matter networks and offers a pathway towards closed-loop material discovery: simulations generate mechanistic datasets, machine learning identifies optimal and high-uncertainty regions of parameter space, and targeted experiments refine and validate predictions. Such iterative workflows are anticipated to be essential for designing programmable soft-matter networks with application-specific mechanical performance in the future.

\section*{Supplementary information}

Supplementary information is available, including detailed model definitions, parameter mappings and additional structural and kinetic analyses.

\section*{Acknowledgements}

IDS and PF gratefully acknowledge funding from the Carl-Zeiss-Stiftung (Center SynGen). IDS further acknowledges the Karlsruhe Institute of Technology Excellence Strategy via the Young Investigator Group Preparation Program. All authors gratefully acknowledge the computational resources provided by the laboratory of Moritz Kreysing from Karlsruhe Institute of Technology (KIT), and thank Aaron Gadzekpo (KIT) for support with data analysis. The authors employed ChatGPT5 for the purposes of proofreading and language improvement.

\section*{Author contributions}

MHS and ACM performed the majority of the simulations featuring in this manuscript. EL and CN provided guidance to MHS and ACM with the execution and interpretation of the simulation results. PF outlined the strategy to be used for active learning and advised ACM on the steps to follow in all results involving machine learning. IDS supervised the work, acquired third-party funding for this research and together with MHS, ACM and EL wrote the first draft. All authors contributed to subsequent revisions of the manuscript and approved the final version.

\section*{Data and code availability statement}
Data and code available from the corresponding author upon reasonable request. 

\section*{Conflict of interests}
The authors declare no conflict of interests. 

\printbibliography
\newpage
\end{document}